\documentclass[11pt,letterpaper]{article}
\usepackage[T1]{fontenc}
\usepackage[utf8]{inputenc}
\usepackage{newtxtext}
\usepackage[margin=0.8in,headheight=14pt]{geometry}
\usepackage{graphicx,booktabs,array,caption,enumitem,titlesec,fancyhdr}
\usepackage{xurl}
\usepackage[hidelinks,unicode]{hyperref}
\titleformat{\section}{\large\bfseries}{\thesection.}{0.5em}{}
\titleformat{\subsection}{\normalsize\bfseries}{\thesubsection.}{0.5em}{}
\titlespacing*{\section}{0pt}{12pt}{5pt}
\titlespacing*{\subsection}{0pt}{9pt}{4pt}
\setlist[itemize]{topsep=3pt,itemsep=1pt,parsep=0pt,leftmargin=1.7em}
\hypersetup{pdftitle={Where Does Streaming State Cost Go? A Reproducible Comparison of Flink and Kafka Streams on Kafka},pdfauthor={Kiran N Kumar; Santhosh Kumar Saminathan}}
\begin{document}

\begin{center}

{\normalsize \textit{Original Article}\par}

\vspace{4pt}

{\LARGE\bfseries Where Does Streaming State Cost Go? A Reproducible Comparison of Flink and Kafka Streams on Kafka\par}

\vspace{4pt}

{\large Kiran N Kumar\textsuperscript{1}, Santhosh Kumar Saminathan\textsuperscript{2}\par}

\vspace{4pt}

{\normalsize \textsuperscript{1}Indiana University, Bloomington, IN, USA\par}

\vspace{4pt}

{\normalsize \textsuperscript{2}Independent Researcher, CA, USA\par}

\vspace{4pt}

{\small \textsuperscript{1}Corresponding Author: knkumar@iu.edu\par}

\vspace{4pt}

\end{center}

Received: Revised: Accepted: Published:

\medskip

\textbf{Abstract} This paper presents a controlled comparison of exactly-once Kafka pipelines implemented with Apache Flink and Kafka Streams,  two engines with different state-management architectures. The state management in Flink occurs through checkpoints in external storage whereas Kafka Streams restores local state by replaying broker changelogs. The experiments evaluate the effects of the two designs on latency, resource consumption, configuration sensitivity, and recover from injected failures. The engines produced expected outputs across 50 correctness trials, five for each engine-workload combination (exact 95\% CI 0.929-1.000). The latencies however showed differences when tested in 30 minute trials at a fixed rate of 100 events/sec. The measured ingestion to output interval was 4-6 ms for stateless workloads and 4.3-6.2 s for windowed workloads. The median p99 was 1.97-1.97s for stateless workloads and 5.68-9.92s for windowed respectively.  Further, a sensitivity study showed that increasing the durability interval from 1,000 to 10,000 ms shifted the latency distributions in Kafka Streams and Flink between stages. Kafka Streams’ W3 median p99 T2-T1 increased from 6,085 to 23,155 ms; Flink’s T3-T2 increased from 732 to 8,702 ms. Failure injection experiments demonstrate that latency can interfere with incomplete processing behavior. The results show that exactly-once correctness is a necessary but insufficient measure of stream processing performance. Kafka Streams completed 0/5 stateful JVM-kill and 1/5 in each stateful local-volume-loss trials, while corresponding Flink cells completed 5/5. The cost placement and magnitude depends on state management, architecture, configuration, and measurement location. Accordingly, evaluations of fault-tolerant streaming systems should holistically measure latency, completion, recovery behavior, and correctness.

\medskip


\section{Introduction}

Streaming engines are being used in mainstream production systems where data should be processed reliably and efficiently with minimal delay. Various benchmarks on streaming engines typically report throughput, latency, and resource use. These metrics highlight the system performance during a run, but do not reveal bottlenecks associated with state management and recovery. Two engines can process the same Kafka workload and produce same records performing durability and recovery work at different stages. The resulting delay could appear as checkpoint traffic, changelog replay, event waiting, state restoration, or delayed downstream visibility which is not captured in aggregate metrics.

Apache Flink and Kafka Streams particularly make an interesting comparison because they handle state management differently. Flink makes use of distributed checkpoints and restores job state from external storage [1], [2]. On the other hand, Kafka Streams maintain local state in the application and restore it by replaying the broker changelogs [3], [4]. Although the two systems support exactly-once Kafka processing their designs expose different failure and recovery paths. Aggregate metrics such as total throughput and latency cannot reasonably differentiate these paths.

Earlier streaming benchmark studies have compared engines, workloads, selectivity, memory use, and scalability [9]-[13]. They provide useful performance baselines, but generally report how much delay occurs without showing where it arises in the streaming pipeline. This study addresses the measurement gap by using a method to track processing at three points outlined below. Further, it also checks for correctness of the output and records how the system behaves under controlled failure conditions.

The system’s processing path is split into three intervals allowing measurement of time an event spends in each stage, outlined below: 

\begin{itemize}

\item before broker ingestion, 

\item before engine output, and 

\item after engine output. 

\end{itemize}

The advantage of this split is that it isolates where time is spent in the pipeline. However, it does not identify a specific cause for each delay. For example, in windowed workloads that produce only final results, the measured time includes waiting for events as well as processing the data. Such an analysis is out of the scope of this study and is left for future research.

The question here is not which engine is universally faster. It is where observable delay appears under the same deterministic input and output contract, and whether the pipeline completes with correct final output after selected faults

\section{Background and architectural motivation}

The following are a few earlier studies which have examined how stream processing behaves under different workloads and deployment configurations. Nexmark uses continuous auction queries to evaluate stream processing engines [9]. The Yahoo Streaming Benchmark compared Storm, Flink, and Spark streaming across representative applications [10]. ESPBench evaluated enterprise style stream workloads over Kafka [11]. Theodolite focused on scalability in distributed stream-processing deployments [12], and DSPBench describes workload characteristics such as selectivity, processing cost, and memory use [13]. A review of the studies demonstrates how workload choice and system scale influence comparisons between streaming engines.

In addition, Karimov et al. separate the driver from the system under test and define sustainable throughput and latency for windowed operators [17]. ShuffleBench supplies a configurable benchmark for shuffling workloads and compares throughput, latency, and scalability [14]. Henning and Hasselbring focus on a reproducible study of stream-processing frameworks deployed as microservices [16]. Vogel et al. directly benchmark fault recovery across Flink, Kafka Streams, and Spark Structured Streaming [15]. The studies above provide rigorous aggregate and recovery baselines. The scope of this present study is much narrower. Given the same workload rates, where does delay occur and how does state management architecture affect the delay?

\subsection{Motivation}

 Aggregate job latencies could determine if one run is slower compared to another. However, it cannot show if the difference occurs before the broker accepts an input, while the engine is processing it, or after the engine has produced output. Adding a resource trace could reveal increased I/O activity but would still not reveal a source for the delay.

Flink and Kafka streams architectures also differ in how they recover state. The two systems use RocksDB internally for state management to support exactly-once Kafka processing. As noted earlier, Flink uses checkpoints and Kafka Streams uses a compacted changelog. Kafka deployment in this study uses KRaft for controller coordination and broker metadata management [5]. The recovery mechanisms are part of the architecture and not interchangeable in the systems under study. Recovering from a process failure would require each application to handle local state restoration through their provided mechanisms. Further, a broker or storage failure may affect the recovery process and event windows can delay output when the engine is healthy. As a result, output correctness does not imply similar latency or recovery behavior across the two engines.

Taking into account these architectural differences, the experimental harness treats correctness and delay as separate observations. Correctness of the final output is checked by a multiset verifier and a timestamp chain measures delay across the three intervals in the pipeline. The associated CPU, memory, network, and block I/O activity are recorded through resource sampling for analysis. Failure injection tests are reported (within an observation window) and evaluated by completion and visible outcomes. A trial that does not complete within this window is treated as a censored observation and will not be considered as proof of permanent failure or data loss. 

The experiments in this study provide a basis for comparing the two architectures and understanding their behavior beyond a single aggregate number. 

\subsection{Key Terms}

\begin{itemize}

\item State is information retained across records, such as a keyed count or window sum. 

\item A checkpoint is a coordinated Flink snapshot used as a restart point [1]. 

\item A changelog is a Kafka topic that records state-store updates for restoration [3]. 

\item Exactly-once processing means that the tested transaction and isolation settings expose each committed logical result once to the observer; it is evaluated here by final multiset equality, not assumed from configuration. 

\item Final-only output suppresses intermediate updates until a window closes. 

\item Visibility delay is T3-T2, from the engine\textquotesingle{}s output timestamp to observation by a read\_committed consumer. 

\item A right-censored trial did not meet the completion gate within the observation window; DNF means that no conditional delay can be estimated because no trial in that cell completed.

\end{itemize}

\subsection{Research questions}

The experiment primarily examines four questions: 

RQ1: What final correctness and observable latency are recorded for stateless and stateful workloads under the common pipeline?

RQ2: What CPU, memory, network, and block-I/O footprint is present in the archived container traces, and what comparisons do those traces not support?

RQ3: How sensitive are observed T0-T3 intervals to Flink checkpoint and Kafka Streams commit intervals?

RQ4: What completion and final-correctness outcomes occur under JVM termination, broker interruption, and engine-container/local-volume loss?

These questions are evaluated with experiments streaming at 100 events/second across five workloads, using fixed software versions, and a single shared host. The main contribution of this study is a reproducible experimental harness that allows the measurements to understand where the operational cost of state management appears.

\section{Method}

This section outlines the methodology, focusing on controlled workloads, timestamp-based latency measurement, resource sampling, configuration changes, and failure injection tests. Each component isolates an aspect of system behavior while keeping the workload rate, execution environment, and measurement procedure similar across the two engines.

\subsection{Controlled Workloads}

The experiment first varies the workload complexity while keeping the surrounding pipeline fixed. The five workloads vary from stateless processing to stateful stream-stream joins:

\begin{itemize}

\item W1 identity: a stateless identity transformation.

\item W2 filter/map: a deterministic parity filter followed by a doubling operation.

\item W3 tumbling count: events grouped by key in 60-second windows.

\item W4 sliding sum: values summed by key over a 10-minute window with a 1-minute slide.

\item W5 stream-stream join: two input streams intersected within a 10-minute join window.

\end{itemize}

The workload progression affects both the expected output and the timing behavior of that output. W1 and W2 are stateless controls. W3 and W4 exercise keyed, final-only window state under the same observer contract. It is important to note that W3 and W4 emit results only when their windows close, so their output is delayed by design. W5 is included as a bounded join since its output can grow exponentially as the number of matching records increases. This sequence was chosen to isolate the effects of selectivity, keyed window state, overlapping windows, and two-stream correlation while keeping the surrounding Kafka path fixed. To ensure consistency across trials all workloads use the same 100 keys and random seed (7). The multiset verifier, indicated earlier, compares results with expected output and records any missing, unexpected, or duplicate records. Verification in this context depends on content rather than sequence; therefore, output order is not used in any analysis.

\subsection{Timestamp decomposition}

Relying on the defined workload behavior, the experiment then measures delay in processing a record along the pipeline. Each record is timestamped at four points as it moves through the pipeline, and deterministic identifiers allow it to be matched to an originating record. The timestamp generator writes T0 immediately before it submits the record to Kafka, then next the broker’s LogAppendTime provides T1. A timestamp, T2, is recorded by the engine when it produces an output. Finally, T3 is recorded by a consumer when the output is visible.

The timestamps define three intervals on the pipeline:

\begin{itemize}

\item T1-T0: interval for record generation to ingestion

\item T2-T1: interval for ingestion to output

\item T3-T2: interval for output to downstream visibility

\end{itemize}

The calculated intervals indicate where delay appears in the pipeline but cannot identify the internal operation responsible for it. This distinction is particularly relevant for workloads W3 and W4 because the T2-T1 interval includes event-time progression and time spent waiting for a window to close and trigger, in addition to record processing time. The harness does not capture the time at which a window first becomes eligible to emit (Te). Consequently, T2-T1 is reported as the observed processing path delay rather than as isolated engine computation time. 

Figures 1 and 2 illustrate this arrangement. Figure 1 shows the architecture of the experimental harness and Figure 2 shows the sequence of timestamps for a single record.

\begin{figure}[!htbp]
\centering
\includegraphics[width=\linewidth]{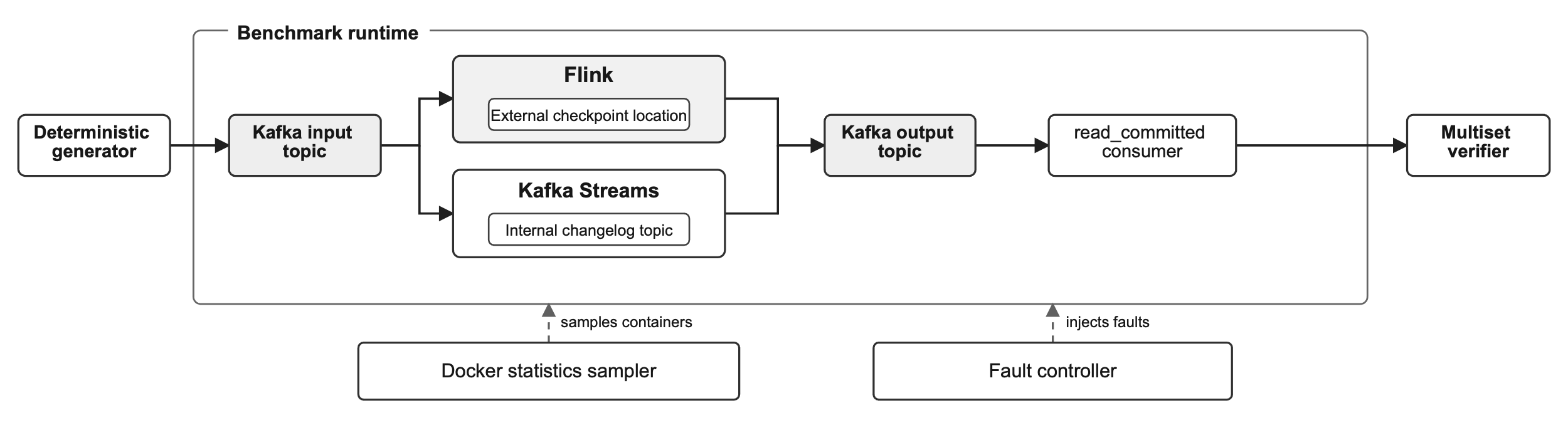}
\caption*{Figure 1 Experimental harness architecture }

\end{figure}
\begin{figure}[!htbp]
\centering
\includegraphics[width=\linewidth]{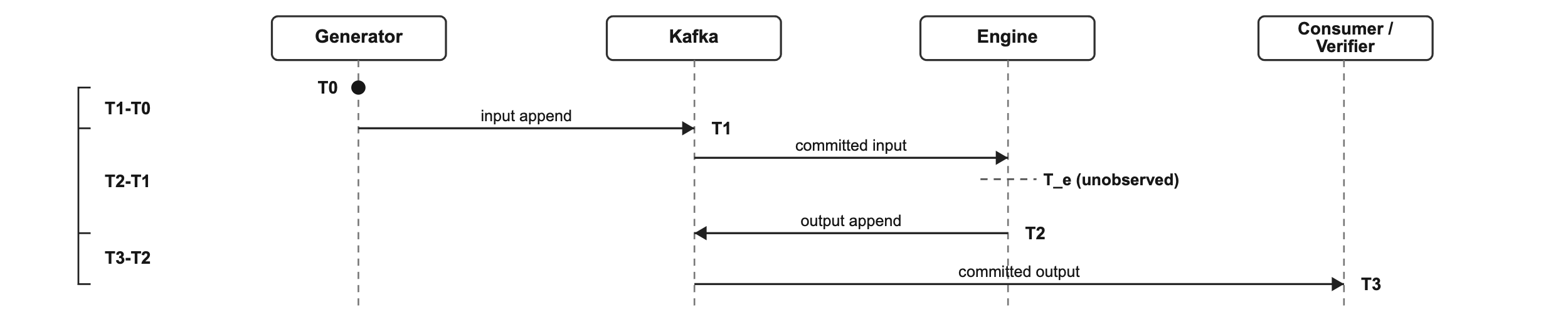}
\caption*{Figure 2 Sequence for timestamp generation and intervals}

\end{figure}

\subsection{Environment setup}

The timestamp procedure is evaluated in a common environment. Both engines use equivalent three-broker apache/kafka:4.3.1 deployments in KRaft mode, and the same external JSONL verifier checks their output. Kafka Streams runs with exactly\_once\_v2. Flink 2.2.0 writes through an exactly\_once sink using flink-connector-kafka 5.0.0-2.2. The connector documents support for Flink 2.1.x and 2.2.x, but not 2.3.x, hence the experiment uses Flink 2.2.0.

The environment shares a single Docker daemon, CPU, and disk, reducing differences from separate hardware environments but sacrificing resource isolation. However, since engines are not assigned to dedicated hosts and identical resource limits are not enforced, the resource figures reported are observations from a shared host rather than isolated hardware costs. To allow a fair comparison, both engines run separately with the same generated input, rate, topic settings, timestamp probe, read\_committed observer, and output verifier. All three broker containers and the engine run on one physical host. The benchmark input and output topics each use one partition, Kafka Streams uses one stream thread, Flink’s operator parallelism is one, and container traffic remains in local networks. 

In this environment, the experimental harness reports five trials per engine-workload for correctness validation, three trials per engine-interval sweep for configuration sensitivity analysis, and five trials each for failure injection conditions. Exhaustive experimentation evaluating system saturation, partitioning, parallelism, and state size is outside the scope of this study. Instead, the focus remains on producing repeatable measurements of latency at each stage, output correctness, resource monitoring, and recovery under fixed control conditions.

\section{Results}

\subsection{Correctness and latency}

The randomized matrix contains 50 executions, five trials for each combination of the two engines and five workloads. All trials in the experiment matched their expected output, with no missing, duplicate, or unexpected records. The number of output records from trials ranged from 199 for W3 to 11,024 for W5. These results establish correctness under the read\_committed configurations.

\begin{table}[!htbp]
\centering
\small
\caption*{Table 1 Decomposed p99 latency in milliseconds under fixed load 100 events/sec. Each row reports the median with CI intervals in square brackets.}

\setlength{\tabcolsep}{4pt}
\renewcommand{\arraystretch}{1.25}
\begin{tabular}{>{\raggedright\arraybackslash}p{0.2\dimexpr\linewidth-10\tabcolsep\relax}>{\raggedright\arraybackslash}p{0.2\dimexpr\linewidth-10\tabcolsep\relax}>{\raggedright\arraybackslash}p{0.2\dimexpr\linewidth-10\tabcolsep\relax}>{\raggedright\arraybackslash}p{0.2\dimexpr\linewidth-10\tabcolsep\relax}>{\raggedright\arraybackslash}p{0.2\dimexpr\linewidth-10\tabcolsep\relax}}
\toprule
\textbf{\textbf{Engine/Workload}}
 & 
\textbf{\textbf{p99 total }}
 & 
\textbf{\textbf{p99 t1-t0 (ingestion)  }}
 & 
\textbf{\textbf{p99 t2-t1 (engine + event-time wait)  }}
 & 
\textbf{\textbf{p99 t3-t2 (commit + consume)  }}
 \\
\midrule
Kafka Streams/W1
 & 
1973.3 [1953.8,1982.8]
 & 
1000.0 [999.8,1000.6]
 & 
11.0 [10.0,17.0]
 & 
1021.0 [1019.2,1097.6]
 \\
Kafka Streams/W2
 & 
1981.1 [1958.9,1983.1]
 & 
1000.0 [999.9,1000.4]
 & 
12.0 [10.0,15.0]
 & 
1022.5 [1017.9,1107.3
 \\
Kafka Streams/W3
 & 
6715.6 [6605.5,6749.7]
 & 
993.7 [992.3,995.1]
 & 
6195.0 [6083.0,6289.0]
 & 
73.8 [43.2,100.4]
 \\
Kafka Streams/W4
 & 
6919.6 [6829.7,6941.1]
 & 
994.4 [993.6,999.9]
 & 
6270.0 [6267.0,6275.0]
 & 
86.4 [80.5,91.9]
 \\
Flink/W1
 & 
1971.8 [1967.6,1973.3]
 & 
1000.4 [1000.1,1000.8]
 & 
97.0 [90.0,99.0]
 & 
1072.0 [1064.4,1078.6]
 \\
Flink/W2
 & 
1977.2 [1973.9,1979.5]
 & 
1000.4 [1000.2,1000.5]
 & 
93.0 [90.0,99.0]
 & 
1064.9 [1048.0,1079.5]
 \\
Flink/W3
 & 
5678.6 [5634.1,5698.2]
 & 
999.1 [992.2,1000.4]
 & 
4282.0 [4280.0,4292.0]
 & 
1025.6 [1019.0,1084.6]
 \\
Flink/W4
 & 
5841.5 [5793.3,5933.5]
 & 
993.2 [991.7,1000.5]
 & 
5108.0 [5099.0,5117.0]
 & 
1083.5 [1064.9,1117.7]
 \\
\bottomrule
\end{tabular}
\end{table}

In the finite input trials, 100 records are processed at 10, 20, and 40 records/second. Across these trials, p99 end-to-end latency was 1.97 to 1.98 seconds for the stateless workloads. For the stateful workloads, it was 6.72 seconds for Kafka Streams and 5.68 seconds for Flink on W3, and on w4, 6.92 seconds and 5.84 seconds respectively. Stateful workloads exhibited approximately 3 times the p99 latency of the stateless workloads.

The fixed load experiments provide a more robust comparison. The fixed load trials use 180,000 inputs, approximately 1,800 times more than finite input trials.

\subsection{Fixed load latency and resource consumption}

Under fixed load conditions, W1-W4 ran for 30 minutes at 100 events/sec and processed 180,000 inputs each. Neither engine showed continuously increasing backlog, and all 
executions passed verification. The verified outputs totaled 180,000 for W1, 89,906 for W2, 29,924 for W3, and 30,900 for W4.


Table 1 reports p99 total latency along with observed timestamp intervals. T1-T0 stayed near one second, but the experimental harness cannot separate producer batching from broker processing. For the stateless workloads W1-W2, the median p99 T2-T1 interval was 11-12 ms for Kafka Streams and 93-97 ms for Flink. For the windowed workloads W3-W4, it increased to 6.20-6.27 seconds for Kafka Streams and 4.28-5.11 seconds for Flink. The larger values observed for Kafka Streams indicate a longer measured interval, but not a proportional increase in computation. As mentioned earlier, W5 was evaluated as a bounded workload and therefore not included in the 30-minute fixed load trials.

Resource monitoring via Docker captured container CPU, active memory, and cumulative network and block I/O. Average Kafka Streams worker memory increased from 117.5 MiB for W1 to 157.0 MiB for W4, whereas Flink averaged 377.6 MiB across trials. Notably, these shared-host values do not represent normalized system costs.

The sensitivity study for W3 examines how durability interval affects the location of observed delays. Kafka Streams varies COMMIT\_INTERVAL\_MS and Flink varies CHECKPOINT\_INTERVAL\_MS. In each of the 12 experiments 6,000 inputs are processed at 100 events/sec for 60 seconds, then event time is advanced to close the windows. Three trials per engine-interval setting each produced 997 expected final-window records. Table 2 gives the median of three nearest-rank p99 values per configuration, bracketed values are confidence intervals. For Kafka Streams, increasing the interval from 1,000 to 10,000 ms increased median p99 total latency from 6,767.1 ms [6582.2,6826.4] to 23908.1 ms [23880.6,23949.6]. T2-T1 increased from 6085 ms [600,6092] to 23155 ms [23046,23187], while T3-T2 changed from 93.7 ms [88.3,124.1] to 137.2 ms [131.5,151.6]. Flink showed a different distribution, median p99 total latency increased from 5596.7 ms [5508.8,5863.6] to 12643.2 ms [12157.5,13509.5]. T2-T1 remained similar at 4259 ms [4236,4271] and 4240 ms [4230–4255], whereas T3-T2 increased from 732.1 ms [693.7,1087.6] to 8702.2 ms [8698.1,10041.3].

\begin{table}[!htbp]
\centering
\small
\caption*{Table 2 Configuration sensitivity of the W3 tumbling-count workload to checkpoint and commit interval. Each row reports the median with CI intervals in square brackets.}

\setlength{\tabcolsep}{4pt}
\renewcommand{\arraystretch}{1.25}
\begin{tabular}{>{\raggedright\arraybackslash}p{0.14\dimexpr\linewidth-12\tabcolsep\relax}>{\raggedright\arraybackslash}p{0.12\dimexpr\linewidth-12\tabcolsep\relax}>{\raggedright\arraybackslash}p{0.185\dimexpr\linewidth-12\tabcolsep\relax}>{\raggedright\arraybackslash}p{0.185\dimexpr\linewidth-12\tabcolsep\relax}>{\raggedright\arraybackslash}p{0.185\dimexpr\linewidth-12\tabcolsep\relax}>{\raggedright\arraybackslash}p{0.185\dimexpr\linewidth-12\tabcolsep\relax}}
\toprule
\textbf{\textbf{Engine}}
 & 
\textbf{\textbf{Interval}}
 & 
\textbf{\textbf{p99 total}}
 & 
\textbf{\textbf{p99 t1-t0}}
 & 
\textbf{\textbf{p99 t2-t1 (engine + event-time wait)}}
 & 
\textbf{\textbf{p99 t3-t2 (commit)}}
 \\
\midrule
Flink
 & 
1000 ms
 & 
5596.7 [5508.8,5863.6]
 & 
1002.0 [1001.8,1002.0]
 & 
4259.0 [4236.0,4271.0]
 & 
732.1 [693.7,1087.6]
 \\
Flink
 & 
10000 ms
 & 
12643.2 [12157.5,13509.5]
 & 
1001.8 [1001.0,1001.8]
 & 
4240.0 [4230.0,4255.0]
 & 
8702.2 [8698.1,10041.3]
 \\
Kafka streams
 & 
1000 ms
 & 
6767.1 [6582.2,6826.4]
 & 
1001.8 [1001.2,1001.9]
 & 
6085.0 [6002.0,6092.0]
 & 
93.7 [88.3,124.1]
 \\
Kafka streams
 & 
10000 ms
 & 
23908.1 [23880.6,23949.6]
 & 
1001.4 [1001.3,1001.5]
 & 
23155.0 [23046.0,23187.0]
 & 
137.2 [131.5,151.6]
 \\
\bottomrule
\end{tabular}
\end{table}

The results indicate that altering the durability interval shifted the stage where delay became observable. The interval shift is evident, but its cause remains undetermined. In Kafka Streams, the T2-T1 includes event-time waiting; and caching was not varied independently of the commit interval. In Flink, downstream visibility also includes checkpoint barriers and transaction boundaries. To isolate these effects, the experimental harness would need to measure stream time, cache behavior, checkpoint barriers, transaction boundaries, and output visibility.

\subsection{Fault injection experiments}

Fault injection experiments were run for W1, W3, and W4 under three conditions:

\begin{itemize}

\item jvm\_kill: termination of the JVM

\item broker\_kill: termination of Kafka broker

\item node\_loss: recreation of engine container after removal of its local state volume

\end{itemize}

All W1 trials matched the expected output. However, for W3-W4 several Kafka Streams jvm\_kill and node\_loss trials did not complete within the fixed window. The corresponding Flink trials completed and passed verification. To ensure a fair comparison, trials that did not complete are treated as censored observations rather than as confirmed failures. The harness telemetry records output timestamps and final verification, but it does not record all recovery milestones. In particular, specific recovery milestones such as the exact failure time, reassignment, state restoration, first output post recovery, or backlog drain milestones are not captured.

Therefore, Table 3 reports the median p99 T2-T1 after fault injection, instead of direct recovery duration. Cells marked DNF indicate fewer than three of the five trials completed, the completion proportion is given in parentheses.

\begin{table}[!htbp]
\centering
\small
\caption*{Table 3 Median per-run p99 latency in seconds T2-T1 after fault injection (Five trials per cell). Each row reports the median with CI intervals in square brackets.}

\setlength{\tabcolsep}{4pt}
\renewcommand{\arraystretch}{1.25}
\begin{tabular}{>{\raggedright\arraybackslash}p{0.16\dimexpr\linewidth-10\tabcolsep\relax}>{\raggedright\arraybackslash}p{0.16\dimexpr\linewidth-10\tabcolsep\relax}>{\raggedright\arraybackslash}p{0.226\dimexpr\linewidth-10\tabcolsep\relax}>{\raggedright\arraybackslash}p{0.227\dimexpr\linewidth-10\tabcolsep\relax}>{\raggedright\arraybackslash}p{0.227\dimexpr\linewidth-10\tabcolsep\relax}}
\toprule
\textbf{\textbf{Engine}}
 & 
\textbf{\textbf{Failure mode}}
 & 
\textbf{\textbf{W1 p99 T2-T1}}
 & 
\textbf{\textbf{W3 p99 T2-T1}}
 & 
\textbf{\textbf{W4 p99 T2-T1}}
 \\
\midrule
Flink
 & 
jvm\_kill
 & 
5/5; 10.19 [10.06,10.66]
 & 
5/5; 21.86 [21.06,22.15]
 & 
5/5; 21.12 [20.99,21.15]
 \\
Flink
 & 
broker\_kill
 & 
5/5; 0.10 [0.08,0.13]
 & 
5/5; 21.17 [21.11,22.27]
 & 
5/5; 21.09 [20.70,21.32]
 \\
Flink
 & 
node\_loss
 & 
5/5; 10.46 [8.46,10.88]
 & 
5/5; 21.21 [21.09,22.17]
 & 
5/5; 21.07 [20.96,21.14]
 \\
Kafka streams
 & 
jvm\_kill
 & 
5/5; 43.55 [43.20,44.13]
 & 
0/5; DNF
 & 
0/5; DNF
 \\
Kafka streams
 & 
broker\_kill
 & 
5/5; 0.07 [0.05,0.11]
 & 
5/5; 26.30 [22.58,32.65]
 & 
5/5; 23.13 [22.86,24.48]
 \\
Kafka streams
 & 
node\_loss
 & 
5/5; 42.94 [41.66,43.16]
 & 
1/5; 56.85 [single run]
 & 
1/5; 49.48 [single run]
 \\
\bottomrule
\end{tabular}
\end{table}

Both engines successfully completed every broker\_kill trial. For W1, the conditional median of five per-run p99 T2–T1 values was 99 ms for Flink, with a CI of 84-129 ms, and 67 ms for Kafka Streams, with a CI of 49-112 ms. The similarity indicates that the broker interruption affected both engines equally because they rely on the shared Kafka deployment.

All trials reported in stateful Flink cells completed, whereas several Kafka Streams W3-W4 cells under jvm\_kill and node\_loss did not. These results establish a difference in observed completion behavior under the tested conditions, but they do not identify the specific recovery mechanism responsible.

The fault experiments underscore the inadequacy of evaluating recovery using latency alone. A comprehensive evaluation must also consider whether processing resumes, whether the backlog effectively drains, expected outputs eventually become visible, and final outputs correctness is maintained.

\section{Discussion and Scope}

The results indicate that correctness and delay characterize two very different sides of system behavior. Although both engines consistently produced accurate outputs across all our trials, latency varied depending on the workload. The processing time (T2-T1) was fast for stateless workloads but increased significantly for windowed workloads. The interpretation needs caution, since the time for W3 and W4 includes event-time progression, window triggering, and processing. Ultimately, this study does not determine if Flink or Kafka Streams is universally faster.

The configuration sensitivity study reinforces this distinction. Increasing the durability interval affected different stages of the pipeline in the two systems. For Kafka Streams, noticeable change appeared in processing T2-T1, whereas for Flink it appeared in consumption T3-T2. The timestamp decomposition shows where the additional delay is visible, but the experimental harness cannot identify internal operations that cause it. Despite this limitation, locating the affected interval provides greater interpretability than relying on a single aggregate value. It demonstrates that the costs associated with state durability may appear at various stages, including processing, checkpointing, transaction completion, changelog restoration, or downstream output visibility.

Furthermore, the failure injection experiments revealed a distinct operational difference between the two architectures. While all broker-kill trials completed successfully across both engines several Kafka Streams windowed trials did not complete within the observation window following JVM termination or local state loss. Although these outcomes do not conclusively establish permanent data loss or a weakness in either recovery mechanism, they underscore that relying exclusively on steady-state throughput obscures critical variations in post-failure recovery. Consequently, comprehensive evaluations of stateful streaming systems must analyze the entire process of maintaining and recovering state, instead of treating state overhead merely as a black-box contribution to total latency.

\subsection{Measurement Validity}

The primary measurement limitation concerns the interpretation of T2-T1 interval in windowed workloads, this conflates event-time waiting and window triggering, with operator processing. The interval is reported as observed processing path delay, not isolated computation time, because the harness does not record the emission eligibility milestones.

A second limitation pertains to fault outcomes. The experimental harness records trial completion, output correctness, and latency but omits granular recovery metrics such as state restoration duration or time to drain the backlog. Consequently, trials failing to complete within the observation window are classified as incomplete, censored observations, rather than definitive evidence of permanent failure or data loss.

\subsection{Scope of the Study}

To ensure a reliable comparison, the experimental scope is intentionally controlled. The harness uses a fixed input rate, five workloads, fixed software versions, and a single shared host. The design minimizes variation in workload generation, hardware, and measurement procedures providing a baseline for evaluation.

Consequently, the results should not be generalized beyond this setup. The experiments do not estimate maximum sustainable throughput, identify saturation points, or exhaustively vary partition count, parallelism, state size, cluster size, or storage architectures. It is worth noting that a different deployment environment could produce different latency, resource, and recovery patterns.

The study should be viewed as a controlled examination of where delay and incomplete recovery are observable, rather than as a comprehensive benchmark of the full capacity of either engine.

\section{Conclusion and Future Work}

This study demonstrates that the operational cost of state management is better understood as a progression through the streaming system than as a single metric. The experiments revealed that delays materialize at fundamentally different stages depending on the engine\textquotesingle{}s architecture and the workload\textquotesingle{}s semantics.

Furthermore, the configuration and fault-injection experiments highlighted that system behavior under stress cannot be captured by throughput alone. Adjusting durability settings shifted exactly where delays became visible within the pipeline, and the two systems exhibited distinctly different completion behaviors when faced with sudden process or state failures.

Together, these findings suggest that streaming engine evaluations should combine output correctness, stage level timestamps, resource measurements, and bounded recovery outcomes. Throughput remains useful but cannot by itself explain where state-related costs occur or whether processing successfully completes after a failure.

Across 50 correctness trials, both engines produced correct expected output. However, windowed workloads had higher p99 latency than stateless controls and durability settings shifted delay to different measured stages. Under fault injection, several Kafka Streams did not complete within the observation window while the corresponding Flink trials did. In particular, T2-T1 includes event-time waiting and the fault harness does not measure complete recovery timelines. The main takeaway is that exactly once correctness does not always establish comparable latency or recovery behavior. Hence, evaluations should combine output verification, latency at each stage, and fault completion tests on representative stateful workloads for an useful comparison.

Future work should expand the harness to record stream time progression, window eligibility, checkpoint and changelog progress, partition reassignment, state restoration milestones, the first post recovery output, and backlog-drain completion. These signals would help determine costs associated with waiting, computation, durability, recovery, and visibility while preserving the controlled comparison methodology established in this study.

\section{Author Contributions}

KNN conducted the W3, W4, and W5 experiments. SKS conducted the W1 and W2 experiments.

Source repository: \url{https://github.com/knkumar/flink-kafka} 

\section{Conflicts of Interest}

The authors declare that there is no conflict of interest regarding the publication of this paper.

\section{Funding Statement}

This research received no specific grant from any funding agency in the public, commercial, or not-for-profit sectors.

\section{Data and Code Availability}

The source code and experimental artifacts supporting this study are publicly available in the project repository (https://github.com/knkumar/flink-kafka).

\section{References}

\begin{enumerate}[label={[\arabic*]},leftmargin=2.3em,itemsep=5pt]\small\raggedright

\item Apache Flink, “Checkpointing”: \url{https://nightlies.apache.org/flink/flink-docs-stable/docs/dev/datastream/fault-tolerance/checkpointing/}

\item Apache Flink, “Fault Tolerance”: \url{https://nightlies.apache.org/flink/flink-docs-stable/docs/learn-flink/fault_tolerance/}

\item Apache Kafka, “Managing Streams Application Topics”: \url{https://kafka.apache.org/40/streams/developer-guide/manage-topics/}

\item Apache Kafka, “Architecture”: \url{https://kafka.apache.org/31/streams/architecture/}

\item Apache Kafka, “KRaft”: \url{https://kafka.apache.org/35/operations/kraft/}

\item Apache Kafka, “Quickstart”: \url{https://kafka.apache.org/quickstart/}

\item Apache Flink, “Downloads”: \url{https://flink.apache.org/downloads/}

\item Maven Central, org.apache.flink:flink-connector-kafka: \url{https://central.sonatype.com/artifact/org.apache.flink/flink-connector-kafka}

\item P. Tucker, K. Tufte, V. Papadimos, and D. Maier, “NEXMark: A Benchmark for Queries over Data Streams (Draft),” OGI School of Science and Engineering, Oregon Health and Science University, technical report, 2002: \url{https://datalab.cs.pdx.edu/niagara/pstream/nexmark.pdf}

\item S. Chintapalli, D. Dagit, B. Evans, R. Farivar, T. Graves, M. Holderbaugh, Z. Liu, K. Nusbaum, K. Patil, B. J. Peng, and P. Poulosky, “Benchmarking Streaming Computation Engines: Storm, Flink and Spark Streaming,” in Proc. IEEE International Parallel and Distributed Processing Symposium Workshops (IPDPSW), 2016, pp. 1789-1792. doi:10.1109/IPDPSW.2016.138

\item G. Hesse, C. Matthies, M. Perscheid, M. Uflacker, and H. Plattner, “ESPBench: The Enterprise Stream Processing Benchmark,” in Proc. ACM/SPEC International Conference on Performance Engineering (ICPE), 2021, pp. 201-212. doi:10.1145/3427921.3450242

\item S. Henning and W. Hasselbring, “Theodolite: Scalability Benchmarking of Distributed Stream Processing Engines in Microservice Architectures,” Big Data Research, vol. 25, article 100209, 2021. doi:10.1016/j.bdr.2021.100209

\item M. V. Bordin, D. Griebler, G. Mencagli, C. F. R. Geyer, and L. G. L. Fernandes, “DSPBench: A Suite of Benchmark Applications for Distributed Data Stream Processing Systems,” IEEE Access, vol. 8, pp. 222900-222917, 2020. doi:10.1109/ACCESS.2020.3043

\item S. Henning, A. Vogel, M. Leichtfried, O. Ertl, and R. Rabiser, \textquotedbl{}ShuffleBench: A benchmark for large-scale data shuffling operations with distributed stream processing frameworks,\textquotedbl{} in Proc. 15th ACM/SPEC Int. Conf. Performance Engineering, 2024, pp. 2-13, doi: 10.1145/3629526.3645036

\item Vogel, S. Henning, E. Perez-Wohlfeil, O. Ertl, and R. Rabiser, \textquotedbl{}A comprehensive benchmarking analysis of fault recovery in stream processing frameworks,\textquotedbl{} in Proc. 18th ACM Int. Conf. Distributed and Event-Based Systems, 2024, pp. 171-182, doi: 10.1145/3629104.3666040

\item S. Henning and W. Hasselbring, \textquotedbl{}Benchmarking scalability of stream processing frameworks deployed as microservices in the cloud,\textquotedbl{} J. Syst. Softw., vol. 208, art. 111879, 2024, doi: 10.1016/j.jss.2023.111879

\item J. Karimov, T. Rabl, A. Katsifodimos, R. Samarev, H. Heiskanen, and V. Markl, \textquotedbl{}Benchmarking distributed stream data processing systems,\textquotedbl{} in Proc. IEEE 34th Int. Conf. Data Engineering, 2018, pp. 1507-1518, doi: 10.1109/ICDE.2018.00169

\end{enumerate}

\end{document}